\documentstyle[preprint,pre,aps,eqsecnum]{revtex}

\begin{document}
\draft
 
\title{
Quantum Fluctuations Driven 
Orientational Disordering: \\
A Finite-Size Scaling Study }
 
\author{R. Marto\v{n}\'ak$^{(1,2,}$\cite{byline}$^)$, 
D. Marx$^{(3)}$, and P. Nielaba$^{(1)}$}
 
\address{
$^{(1)}$ Institut f\"ur Physik, KoMa 331, Johannes Gutenberg--Universit\"at \\
Staudingerweg~7, 55099~Mainz, Germany \\
$^{(2)}$ Max--Planck--Institut f\"ur Polymerforschung \\
Ackermannweg~10, 55021~Mainz, Germany \\
$^{(3)}$ Max--Planck--Institut f\"ur Festk\"orperforschung \\
Heisenbergstr.~1, 70569~Stuttgart, Germany 
}
 
\date{\today}
\maketitle
 
\begin{abstract}
The orientational ordering transition is investigated in
the quantum generalization of the anisotropic-planar-rotor model
in the low temperature regime.
The phase diagram of the model is first analyzed within the 
mean-field approximation. This predicts at $T=0$ a phase transition
from the ordered to the disordered state when the strength of quantum
fluctuations, characterized by the rotational constant $\Theta$, 
exceeds a critical value $\Theta_{\rm c}^{MF}$. As a function of temperature, 
mean-field theory predicts a range of values of $\Theta$ where the
system develops long-range order upon cooling, but enters again into a
disordered state at sufficiently low temperatures (reentrance).
The model is further studied by means of path integral Monte Carlo
simulations in combination with finite-size scaling techniques, concentrating 
on the region of parameter space where reentrance is predicted to occur. 
The phase diagram determined from the simulations does not seem to exhibit 
reentrant behavior; at intermediate temperatures a pronounced increase 
of short-range order is observed rather than a genuine long-range order.

\end{abstract}

 
\narrowtext

 
%
 
%
%
 
\section{Motivation}
\label{intro}

Physisorbates are experimental realizations of 
quasi two-dimensional systems that display an extremely rich
phase behavior due to the competition between intermolecular
and molecule-surface interactions, as documented, e.g., 
in Refs.~\cite{sinha,taub,wyatt,wiechert,marx1}. 
Correspondingly there is a wealth of phase transitions
between the various ordered phases as a function of temperature
and coverage. 
Since many of these transitions occur at fairly low temperatures,
quantum effects might play an important 
or even crucial role~\cite{wyatt,wiechert}, as recently demonstrated
for the ordering of
hydrogen isotopes on graphite~\cite{kreer}. 
Molecular systems are particularly interesting as they
possess orientational degrees of freedom which can 
order in addition to the positions~\cite{marx1}. 
In case of linear molecules, the  
anisotropic-planar-rotor (APR) model~\cite{aprklein,aprmouritsen} 
was devised to describe the 
herringbone (quadrupolar 
orientational two-sublattice) ordering transition~\cite{marx1},  
e.g., in commensurate N$_2$ monolayers on graphite. 
The classical two-dimensional APR model consists of planar rotators pinned
with their center of rotation on a triangular lattice 
and interacting via nearest-neighbor quadrupolar 
interactions only; a three-dimensional version has also been
proposed and investigated in various approximations~\cite{harris}. 

Over the years the APR Hamiltonian acquired 
the status of a statistical mechanical {\em model} on its own right,
see Refs.~\cite{mouritsen-book,marx1}.
Many of these activities arose because the order of the APR phase transition
turned out to be 
extremely challenging to determine~\cite{schick},
finally favoring a first order 
phase transition that is \lq \lq weak\rq \rq {} and
fluctuation driven~\cite{fo}, see the detailed review in Ref.~\cite{marx1}.
The plain APR model was generalized to include vacancies
or impurities~\cite{vacancies,choi89-doped,mederos90},  
as well as  multipole interactions other than 
those of quadrupolar symmetry~\cite{choi89-doped,choi89}.
Simplified Hamiltonians were obtained after discretizing the continuous 
rotations to an anisotropic 6-state model~\cite{tarazona89}
or to more general discrete models~\cite{vollmayr92}.
We introduced quantum generalizations of interacting
two-dimensional quadrupolar lattice systems~\cite{qn2,qco}, 
in particular    
of the APR model~\cite{qapr0,qapr1}.
For a given lattice, the classical APR model has no free parameter 
because the quadrupolar coupling constant $K$ only sets the 
energy scale of the model.  
Correspondingly, the phase diagram is one-dimensional and
is fully characterized by a single number, 
the transition temperature scaled by the constant $K$. 
In the quantum case, however, the quantum kinetic energy
is determined by 
the mechanical moment of inertia $I$ associated
with the angular motion of the two-dimensional rotators.
The resulting rotational constant $\Theta=\hbar^2/2I$ 
is an additional independent parameter that 
determines the strength and energy scale of the quantum fluctuations,
and in the limit $\Theta \to 0$ the quantum APR model reduces to the classical 
one. 
The resulting behavior of the quantum APR model is 
governed by the interplay between thermal and quantum fluctuations.  
This interesting feature is present also in other systems, such as, e.g., 
Ising models in transverse field~\cite{degennes,stratt}, 
models for granular superconductors~\cite{fazekas} and superconducting 
arrays~\cite{novotny}, lattice $\phi^4$-theory~\cite{jim}, and
quantum four-state clock models~\cite{roman},
see Ref.~\cite{qapr1} for a short discussion of these related models.

In our previous studies~\cite{qapr0,qapr1} 
we presented
the first {\em qualitative} exploration of the 2D quantum APR Hamiltonian,
using path integral Monte Carlo (PIMC) simulations~\cite{pimc}
adapted to 
rotational motion restricted to two dimensions~\cite{rotpimc}.
It was demonstrated numerically~\cite{qapr1} 
that low-order approximation schemes
such as quasiclassical Monte Carlo simulations using
the quadratic Feynman-Hibbs effective potential,
and the simple quasiharmonic approximation
are useful only in the regime
of small rotational constants $\Theta\to 0$, whereas
they fail completely in the large-$\Theta$ range
that is of interest here.
The phase boundaries were estimated phenomenologically from the
behavior of the order parameter, i.e.,
without applying any kind of finite-size scaling to the data
obtained from $30^2 = 900$ interacting rotators.
Increasing the value of $\Theta$, four distinct regimes were found 
based on the shape of the orientational order parameter.
For small $\Theta$, the transition temperature and value
of the ground state order parameter obtained from the classical
model get only renormalized to smaller values. 
In the opposite limit of large $\Theta$, the quantum fluctuations are
that strong that they do not allow for any ordering even in
the ground state. 
More interesting are the intermediate regimes. 
For increasing $\Theta$ a
crossover was found where there is residual ground state order,
although significantly depressed from its classical value, 
but the order parameter inside the ordered phase first {\em increases}  
upon heating and goes through a maximum at some intermediate temperature
before it decays in the disordered phase.
For even larger $\Theta$, there seemed to be a region in 
parameter space with
vanishing ground state order together with
residual order at finite temperatures. 
The \lq \lq tentative and qualitative phase diagram\rq \rq {}
in the $\Theta$--$T$ plane thus 
seemed to exhibit a {\it reentrance} phenomenon in a range of $\Theta$ 
where quantum and thermal fluctuations are competitive,
see Fig.~9 in Ref.~\cite{qapr1}.  
A similar behavior was already discovered in mean-field studies
of certain quantum Ising models~\cite{stratt}, quantum four-state clock models 
with quadrupolar interactions \cite{roman} and  
interacting 3D quadrupolar rotators ~\cite{qmelt}.
The reentrance in the two-level systems
was also found to be present when fluctuations 
were included to lowest order in form of a Kirkwood correction
on top of the Bragg-Williams expressions~\cite{stratt}. 
The reentrance in the three-dimensional system
was ascribed in Ref.~\cite{qmelt} to two distinct 
first order phase transitions,
a standard order-disorder transition at higher temperatures
that is also present in the corresponding classical model, 
and a transition of pure quantum nature at low temperatures. 

With the present paper, we concentrate on the region of the phase diagram
of the 2D quantum APR model
where reentrance might be expected to occur. We use first a mean-field
approximation and then a PIMC simulation. The aim of the work is to study
the orientational transition behavior in the reentrance regime. 
In particular, we are interested to find out whether there 
are actually 
two distinct {\em phase transitions} occurring as a function of temperature.  
To this end, we simulate the system 
on several length scales
and use finite-size scaling, in particular
Binder's cumulant~\cite{blocks,revblocks}, 
to analyze the order parameter distributions. 

The main body of the paper is organized as follows. 
The quantum APR model together with its order parameter
is defined in Sect.~\ref{model}.
In Sect.~\ref{simus}, 
the quantum PIMC simulation technique 
is described, followed by the definition of the relevant observables 
in Sect.~\ref{obs}
and an outline of the finite-size scaling analysis method in 
Sect.~\ref{finitesize}.
The mean-field approximation is worked out in Sect.~\ref{meanfield}. 
In Sect.~\ref{res}, the results of the simulation are discussed and 
compared to those of the
mean-field approximation and a possible scenario is suggested.
In Sect.~\ref{sum}, we summarize the results and draw some conclusions. 
 
\section{Quantum Anisotropic-Planar-Rotor Model}
\label{model}
 
Subject of this study are the properties and in particular
the phase diagram of the
quantum-mechanical version of  
the APR model~\cite{aprklein,aprmouritsen}. 
The classical APR Hamiltonian was extensively studied
over the years, both numerically and analytically, see Ref.~\cite{marx1}
where the properties of the classical model including the
order parameter and the 
herringbone orientationally ordered ground state are discussed. 
The $N$ quadrupolar rotators are fixed with their center of mass on
an ideal rigid triangular lattice $\{ {\bf R}_j \}$ and only rotations in the
two-dimensional surface plane are allowed.
The interactions are truncated at the first-neighbor shell,
and the rotators interact exclusively with
their six nearest neighbors via the anisotropic part of
a quadrupole-quadrupole
potential of strength $K$ in two dimensions.
The quantum generalization results after supplementing this classical 
Hamiltonian with a non-commuting angular
momentum part
$[L_j , \varphi_i ] = - i \hbar \delta_{j,i} $
which introduces quantum dispersion and thus qualitatively new effects
due to additional fluctuations and tunneling.
Correspondingly, the quantum APR Hamiltonian reads
\begin{eqnarray}
H &=&
\sum_{j=1}^N { L^2_j  \over  {2 I} } +
\sum_{ \langle j,i \rangle }^N \> V (\varphi_j, \varphi_i)
\label{ham1}      \\
&=&
- \Theta \sum_{j=1}^N {\partial^2 \over {\partial \varphi_j }^2}
+
K \sum_{ \langle j,i \rangle }^N \>
\cos \left( 2 \varphi_j + 2 \varphi_i - 4 \phi_{j,i}  \right)
\> \> ,
\label{ham2}
\end{eqnarray}
where the angle $\varphi_j$ of the $j$-th rotator
pinned at site ${\bf R}_j$ is 
defined relative to one symmetry axes of the triangular lattice,
and the six phases $\phi_{j,i}$ measure the angle between
neighboring sites ${\bf R}_j$ and ${\bf R}_i$ on this lattice,
i.e., $\phi_{j,i} \in \{ 0, \pi /3, 2\pi / 3, \pi , 4\pi / 3, 5\pi / 3\}$.
We stress that for
this system the triangular lattice structure is essential
in order to produce a nontrivial ordered phase.
On a simple square lattice
the quadrupolar interaction would just favor
unfrustrated perpendicular nearest-neighbor
orientations of the characteristic {\sf T}-shape.

The moment of inertia $I$ determines the
rotational constant $\Theta =\hbar^2 / 2 I$, which is
the parameter that controls the strength of quantum effects.
The other parameter of the model, which is the quadrupolar coupling constant 
$K$, can be conveniently taken as the energy and temperature scale.
We can thus reduce all quantities related to energies by $K$, and define, e.g., 
the dimensionless temperature $T^\star = k_{\rm B} T / K$, energy $E^\star = E/K$,
and rotational constant $\Theta^\star = \Theta / K$.

The long range order parameter sensitive to herringbone ordering of
the rotational axis has three $(\alpha = 1,2,3 )$ 
{\em independent} components
\begin{equation}
\Phi_\alpha \> = \>
{ 1 \over {N} } \sum_{j=1}^N 
\sin ( 2 \varphi_j  - 2 \eta_\alpha )
\exp [ i {\bf Q}_\alpha {\bf R}_j ]
\> \> ,
\label{nop1}
\end{equation}
where
\begin{eqnarray}
{\bf Q}_1 \> = \> & \pi (0,2/ \sqrt{3} ) ,  \> \> \> \>
&\eta_1 \> = \> 0,
  \nonumber  \\
{\bf Q}_2 \> = \> & \pi (-1,-1/ \sqrt{3} ) , \> \> \> \>
&\eta_2 \> = \> 2 \pi / 3,
  \nonumber  \\
{\bf Q}_3 \> = \> & \pi (1,-1/ \sqrt{3} ) ,  \> \> \> \>
&\eta_3 \> = \> 4 \pi / 3.
\label{nopx}
\end{eqnarray}

\section{Numerical Methods}
\label{methods}

\subsection{Path Integral Monte Carlo Simulation Technique}
\label{simus}
 
We study the properties of the Hamiltonian (\ref{ham2}) by means of
PIMC simulations,
for general reviews on the concept of PIMC we refer to Ref.~\cite{pimc}.
We just stress here that
for rotational motion in two dimensions new features connected to
the restricted integration space show up in the formalism~\cite{qrotations}.
This latter aspect
and our implementation of these specialties in a PIMC scheme is discussed
in detail in Ref.~\cite{qn2}, and
here we present only the essential features.

In general the partition function of a Hamiltonian of type (\ref{ham1})
is given by
\begin{eqnarray}
Z  &=&
\text{Tr } \exp \left[ - \beta H  \right]
\label{z1} \\
   &=&
          \lim_{P \to \infty}
          \biggl( {{I P} \over {2 \pi \hbar^2 \beta}} \biggr)^{NP \over 2}
          \prod_{j=1}^{N} \>
          \Biggl\{  \>
          \sum_{n_j =-\infty}^{\infty}
            \> \int\limits_0^{2 \pi} d \varphi^{(1)}_j  \>
          \prod_{s=2}^{P}  \>
          \biggl[\> \int\limits_{-\infty}^{\infty} d \varphi^{(s)}_j \biggr]
          \> \Biggr\}
          \nonumber   \\
          & &
          \times \exp
            \Biggl[
           -\beta  
             {\sum_{s=1}^{P}}
             \Biggl[
             \sum_{j=1}^{N} 
              { {I P} \over {2 \hbar^2 \beta^2} }
              [ \varphi^{(s)}_j - \varphi^{(s+1)}_j
              + 2 n_j \pi \delta_{s,P} ]^2
            +  \sum_{<j,i>}^{N} 
               {1 \over P} V(\varphi^{(s)}_j, \varphi^{(s)}_i)
               \Biggr]    \Biggr]
 \>  \> ,
\label{z3}
\end{eqnarray}
where $V(\varphi^{(s)}_j, \varphi^{(s)}_i)$ denotes
the APR pair potential of (\ref{ham2}) evaluated separately
for the configuration  
at each imaginary time slice $s=1, \cdots , P$. 
Each quantum mechanical rotational degree of freedom is
represented in this path integral representation
by $P$ classical rotators which form closed loops
and interact via harmonic type interactions;
for the related ring-polymer picture see Ref.~\cite{chandler81,pimc}.
The parameter $P$ is the Trotter number, 
the configuration
$\{ \varphi_j^{(1)}, \varphi_j^{(2)}, \cdots , \varphi_j^{(P)} \}$
is a realization of a Trotter path, and the path integral
results from the proper integration and summation over all possible paths. 
However, contrary to path integrals for translational degrees of freedom these
loops need not to be closed using periodic boundary conditions,
but only modulo $2 \pi$; note that the classical angles
$\varphi_j^{(s)}$ of (\ref{z3}) are not confined to
$[ 0 , 2\pi )$ but are allowed on the
whole interval $[ - \infty , \infty ]$.
The resulting mismatch $n_j$ is called the
\lq \lq winding number \rq \rq of
the $j^{\em th}$ path~\cite{qrotations} and the formulation (\ref{z3}) is
the \rq \rq winding number representation\rq \rq {}
of the partition function.
Only the Boltzmann-weighted summation over all possible winding numbers 
in addition to the integration over all paths having a certain
winding number yields the correct quantum partition function in the
Trotter limit $P \rightarrow \infty $;
see Ref.~\cite{qn2} for a full discussion of that issue.
Thus we have to include in the algorithm in addition to
local and global moves of the angular degrees of
freedom $\{\varphi^{(s)}_j\}$ 
also attempts to change the winding numbers $\{n_j\}$ of
the individual rotators. 
Our algorithm was tested~\cite{rotpimc,qn2} against
exact (single-particle) results and
a close agreement was observed.
In a previous study~\cite{qapr1} it turned out that
the use of a finite Trotter number of
$P=500$ at a temperature of $T^\star = 0.03$ and a progressive linear
decrease of $P$ with increasing temperature was sufficient to
be in the large $P$ limit as required in (\ref{z3}).
 
In the present paper we study in particular finite-size effects
on the quantum APR transition. 
The required different linear dimensions $L$ of the system
were chosen to range from  $N$~= $L^2$~= 144 to 900, where 
additional data from Ref.~\cite{qapr1} have been included in
the analysis. 
We equilibrated the systems starting from the ideally
ordered herringbone ground state and made where
possible use of previous runs either at lower temperatures
or smaller rotational constant.
Our statistics is based on the order of 10$^5$ PIMC steps
for each data point, 
a typical CPU time per data point was of the order of 50 hours
on a CRAY--YMP.

\subsection{Observables and Estimators}
\label{obs}
 
The quantities we determined from the simulations are
kinetic and potential energy, order parameter, 
quantum librational amplitudes, as
well as both fourth order (\ref{ul4}) and second order (\ref{ul2}) cumulants.
 
The energies were obtained from the primitive estimator~\cite{pimc}
which proved to be sufficient in the present case~\cite{qn2}.
The kinetic energy $E_{\rm kin}$ and the potential energy $E_{\rm pot}$
per particle
are given by
\begin{equation}
E_{\rm kin} \> = \> {P k_{\rm B} T\over 2} -
                      \bigg\langle
                       {1 \over N}
                       \sum_{j=1}^N
                       \sum_{s=1}^{P}
                        { {I P} \over {2 \hbar^2 \beta^2} }
                        [ \varphi^{(s)}_j - \varphi^{(s+1)}_j
                        + 2 n_j \pi \delta_{s,P} ]^2
                      \bigg\rangle \> \> ,
\label{ekin}
\end{equation}
\begin{equation}
E_{\rm pot} \> = \>
                      \bigg\langle
               {1 \over P}
             {\sum_{s=1}^{P}}
                {1 \over N}
               \sum_{<j,i>}^{N}
               V(\varphi^{(s)}_j, \varphi^{(s)}_i)
                      \bigg\rangle \> \> ,
\label{epot}
\end{equation}
and the total energy $E_{\rm tot}$ is
given by the sum of $E_{\rm kin}$ and $E_{\rm pot}$.
The estimator for the order parameter component $\Phi_\alpha$
is given by the expression
\begin{equation}
\Phi_\alpha \> = \>
{ 1 \over {NP} } \sum_{j=1}^N \sum_{s=1}^P
\sin ( 2 \varphi^{(s)}_j  - 2 \eta_\alpha )
\exp [ i {\bf Q}_\alpha {\bf R}_j ]
\> \> ,
\label{nop2b}
\end{equation}
which trivially follows from (\ref{nop1}).
In the following we will use the total long range order
parameter defined as the length~\cite{qn2,fo}
\begin{equation}
 \Phi  = \left< [ \> \sum_{\alpha=1}^3 \Phi_\alpha^2 \> ]^{1 \over 2 }
\right>
\label{nop2}
\end{equation}
of the three component vector order parameter (\ref{nop2b}).

A quantity which measures the quantum delocalization of the
rotational degrees of freedom
can be defined by the expression~\cite{qn2}
\begin{equation}
R_\varphi \> = \> \bigg\langle
                      {1 \over N}
                      \sum_{j=1}^N
                      {1 \over P}
                      \sum_{s=1}^{P} \Bigl[  \varphi^{(s)}_j -
                                       {1 \over P} \sum_{k=1}^{P}
                                        \varphi^{(k)}_j \Bigr] ^2
       \bigg\rangle ^{1 \over 2}  \> \> .
\label{mspread}
\end{equation}
This average spread is by its definition zero for
a classical system and is thus a measure of the
pure quantum contribution to the librations or rotations
of the individual rotators.
 
\subsection{Finite-Size Scaling}
\label{finitesize}

A fingerprint of a continuous phase transition is the divergence of
the correlation length $\xi$ at the critical temperature $T_{\rm c}$, with 
$\xi \sim |t|^{-\nu}$ for $t:=1-T/T_{\rm c}$, $|t| \ll 1$. 
Under these conditions on
the reduced distance from the phase transition
$t\geq 0$ the order parameter $\Phi$
in the infinitely large system depends on $t$ as $\Phi \sim |t|^\beta$, 
where $\nu$ and $\beta$ denote here the usual critical exponents.
In finite systems of linear dimension $L$ the order parameter
is given by the expression~\cite{revblocks} 
\begin{equation}
\langle \Phi \rangle_L = L^{-\beta/\nu} \tilde{\Phi} (L/\xi)\enspace , 
\label{opl}
\end{equation}
where $\tilde{\Phi}$ is the scaling function that is associated to
$\Phi$.
The order parameter distribution function
$P_L(\Phi)$ in the finite system is
thus a function of the scaling variables $L^{\beta/\nu}\Phi $ and $L/\xi$,
\begin{equation}
P_L(\Phi) = L^{\beta/\nu} \tilde{P}(L^{\beta/\nu} \Phi,L/\xi) \enspace , 
\label{opdis}
\end{equation}
and the $k$-th moment of the order parameter distribution function
is given by
\begin{equation}
\langle \Phi^k \rangle_L = 
      L^{\beta/\nu} \int d\Phi \>
      \Phi^k \tilde{P}(L^{\beta/\nu} \Phi,L/\xi)  
     = 
     L^{-\beta k/\nu} {\tilde\Phi}_k(L/\xi)\enspace , 
\label{opmom}
\end{equation}
where the symbols $\left< \quad \right>_L$ denote
canonical averages with systems of linear dimension $L$.

A very useful quantity for the determination of a critical point 
which is directly
based on order parameter moments is the fourth order 
cumulant~\cite{blocks,revblocks} $U_L^{(4)}$
or the second order cumulant~\cite{HPD} $U_L^{(2)}$
defined as
\begin{equation}
U_L^{(4)} = 1 - {{\left< \Phi^4 \right>_L}  \over
{3 \left< \Phi^2 \right>_L^2 }}
\enspace ,
\label{ul4}
\end{equation}
\begin{equation}
U_L^{(2)} =
1 - {{\left< \Phi^2 \right>_L}  \over {3 \left< \Phi \right>_L^2 }}
\enspace , 
\label{ul2}
\end{equation}
where one can see that the explicit dependence on
system size drops out if the $k$-th moments 
$\sim L^{-\beta k/\nu}$
are re-expressed with the aid of
(\ref{opmom}) in terms of their scaling functions.
We only compile here the main features and usage of these quantities,
and refer for further details to the literature~\cite{blocks,revblocks}.

In the case of second order transitions,
the cumulants adopt a non-trivial universal value 
${U}^\ast$ at the critical
point, irrespective of system sizes $\{ L \} $ in the scaling limit.
Thus, plotting $U_L^{(4)} (T)$ or
$U_L^{(2)} (T)$ for 
different linear dimensions
as a function of temperature yields an intersection point 
$U(T_{\rm c})=:{U}^\ast$ 
which
gives an accurate estimate of the critical temperature
in the {\em infinite} system for a temperature driven second order transition.
Below and above the transition, the cumulants flow to trivial
limiting values depending on the definition of the order parameter;
the larger the system, the faster the 
convergence.
Instead of the temperature, other parameterizations of
approaching a critical point
can also be chosen. 
More recently, it was worked out that the concept of order parameter
cumulant crossings
is also useful to analyze first order phase transitions~\cite{kathrin}. 
In this case, one can observe  
an {\em effective} crossing point
at a non-universal value $U^\ast$ at the phase transition.
The approach of both the transition point and the value of $U$ at the 
transition to the infinite system limit is quite fast:
the correction depends roughly speaking 
on the inverse volume of the system~\cite{kathrin}.  
Thus, for practical numerical purposes the order 
parameter cumulant can be taken as acquiring an intersection point
at a first order transition in a way 
similar to that occurring at a second order transition.  

\section{Mean-Field Theory}
\label{meanfield}

In this section we start from the Hamiltonian (\ref{ham2}) and
determine the phase diagram for our model using a mean-field approximation.
This consists in considering a single quantum rotator in the mean field of its
six nearest neighbors and finding a self-consistent condition for the order 
parameter. Solving the latter condition, we determine the phase boundary and
also the order of the transition. Our mean-field approximation is similar in 
spirit to that used in Ref.~\cite{qmelt} for the case of 3D rotators.

We shall assume that the order parameter component $\Phi_1$ becomes 
non-zero in the ordered phase,
$\Phi_1 = 
\pm \langle \sin 2\varphi \rangle \neq 0$ (we choose positive sign for the single 
rotator, note that the sign alternates passing from one
row of the herringbone structure to another, see Eq.(\ref{nop1})), 
while $\Phi_2 = \Phi_3 = 0$.
Now it is useful for a moment to come back to normal (non-reduced) units for 
$\Theta$, $K$ and $T$.  
In order to proceed, we write each interaction term 
$\enspace K\cos(2\varphi + 2\varphi^{'} - 4 \phi_{ij})$ as a 
product of trigonometric functions depending separately on the angular variables
$\varphi$ of the single rotator and $\varphi^{'}$ of its nearest neighbors. 
Averaging over the variables $\varphi^{'}$ we find the following contributions:
for the two terms corresponding to  $\phi_{ij} = 0$ we get 
$- 2 K \Phi_1 \sin 2 \varphi$, 
for the two terms corresponding to $\phi_{ij} = 
\pi/3$ we get $- K \Phi_1 \sin 2 \varphi + \sqrt{3} K \Phi_1 \cos 2 \varphi$ and
for the two terms corresponding to $\phi_{ij} = 
2\pi/3$ we get $- K \Phi_1 \sin 2 \varphi - \sqrt{3} K \Phi_1 \cos 2 \varphi$.
Summing these contributions from all six nearest neighbors, we find the 
total mean-field potential acting on the single rotator, which reads 
\begin{equation}
H^{\rm MF}_{\rm pot} = - 4 K \Phi_1 \sin 2 \varphi \enspace .
\end{equation}
Adding the kinetic energy, we get the corresponding one-particle
Schr\"{o}dinger equation 
for the on-site problem 
\begin{equation}
H^{\rm MF} \Psi = \left(- \Theta {{d^2}\over{d \varphi^2}} - 
4 K \Phi_1 \sin 2 \varphi \right) \Psi = E \Psi \enspace . \label{schr}
\end{equation}
We now introduce the quantity $q = -2 K \Phi_1 / \Theta$ and perform
a trivial shift of the angular variable $\varphi$ by defining 
$\theta = \varphi - \pi/4$. In terms of these new variables, the eigenvalue
problem (\ref{schr}) can be written as 
\begin{equation}
{{d^2 \Psi}\over{d \theta^2}} + 
({{E}\over{\Theta}} - 2 q \cos 2 \theta)\Psi = 0 \; , \label{mat}
\end{equation}
which is the well-known Mathieu's equation \cite{abram}. Its eigenvalues can
be labeled by an non-negative integer number $m$, and for each $m \neq 0$ there
are two eigenvalues, $a_{m}(q)$ associated with an even periodic solution, and 
$b_{m}(q)$ associated with an odd periodic solution. For $m=0$ there is just 
one eigenvalue $a_{0}(q)$ associated with an even periodic solution.

In order to find the self-consistent condition, we have to determine the
order parameter 
\begin{equation}
\Phi_1 = \langle \cos 2 \theta \rangle = 
{{\sum_i \langle \Psi_i| \cos 2 \theta | \Psi_i \rangle e^{-\beta E_i}} \over
{\sum_i e^{-\beta E_i}}} = {{1}\over{2}} {{\partial f_0} \over {\partial q}} 
\label{op}
\end{equation}
as a function of the parameter $q$. In the last equation we have used the free
energy per site defined as $f_0 = - (1/\beta) \ln \sum_i e^{-\beta E_i}$. 
Solving the
equation (\ref{op}) simultaneously with the condition $q = -2 K \Phi_1/\Theta$,
we find the equilibrium value of the order parameter for given values of inverse
temperature $\beta$ and model parameters $K,\Theta$. 
 
Because our main interest is to 
determine the phase boundary and the order of the 
transition, we do not have to find the complete expression for $\Phi_1(q)$. For
a continuous phase transition, the phase boundary is a curve in parameter space
on which a nontrivial and infinitesimally small solution
$\Phi_1 \neq 0$ appears, apart from the trivial one $\Phi_1 = 0$, which is 
always present and corresponds to the disordered phase. In order to determine
the phase boundary and check the order of the transition, it is enough to
find an expansion of $\Phi_1$ in powers of $q$ up to the third order, which 
can be obtained from the expansion of $f_0$ in powers of $q$ up to the fourth 
order. The expansions of the eigenvalues of the Mathieu equation can be found
in \cite{abram} and we quote here just first few of them, which have the 
following form
\begin{eqnarray}
a_0(q) &=& -{{q^2}\over{2}} + {{7 q^4}\over{128}} - \ldots \\ \nonumber
a_1(-q) &=& b_1(q) = 1 - q - {{q^2}\over{8}} + {{q^3}\over{64}} - 
{{q^4}\over{1536}} \ldots \enspace . \label{eigenvalues} 
\end{eqnarray}

For practical calculations in the low temperature region, we can truncate the 
infinite sum in
$f_0$ and include only a finite number of terms. Increasing the number of terms
would yield a progressively better approximation in the high temperature limit,
but this is actually not necessary, since in this limit the quantum effects 
become negligible and the model behaves classically. The classical model
has already been investigated within various mean-field 
approximations in Ref. \cite{chata}.
We have taken the eigenvalues corresponding to $m \leq 6$. The first neglected 
eigenvalue then 
equals $7^2 = 49$ for $q = 0$ and therefore our approximation should
yield reliable results up to temperatures which are at least an order of 
magnitude lower than the first neglected term, therefore for temperatures 
$T/ \Theta \leq 5$. 
Substituting the energy eigenvalues expansions into (\ref{op}) and
calculating the free energy $f_0$ and order parameter $\Phi_1$ (the actual 
symbolic calculation has been performed by {\em Mathematica}), we find that
the expansion of the latter has the form 
\begin{equation}
\Phi_1 = \chi_1 q + \chi_3 q^3 + \ldots \enspace , \label{opexp} 
\end{equation}
where the coefficients
$\chi_1$ and $\chi_3$ are somewhat complicated expressions. Combining 
the expression (\ref{opexp}) with the equation $q = - 2 K \Phi_1/\Theta$, 
we find that
the condition for the phase boundary is given by
$K_{\rm c} = - \Theta/(2 \chi_1)$. 
We quote here the final result for $K_{\rm c}$, which reads 
\begin{equation}
K_{\rm c} = {{420 \, \Theta_{\rm c}  Z_6} \over {420 +
210 B_1 \left({4 \Theta_{\rm c} \over T_{\rm c}} +1\right)
- 280 B_2 - 105 B_3 - 56 B_4 - 35 B_5 - 24 B_6}} \; , 
\label{phb}
\end{equation}
where $B_m=\exp[-m^2\Theta_{\rm c}/T_{\rm c}]$ and $Z_6 = \sum_{m=-6}^6 B_m$.

We yet have to make sure that the transition is really continuous. The 
necessary and sufficient condition for this is that the coefficient $\chi_3$
is negative on the phase boundary (\ref{phb}). We have found this to be 
always satisfied. Therefore the phase boundary (\ref{phb})
corresponds to a continuous phase transition. In the high-temperature, 
classical limit this agrees with the finding in Ref. \cite{chata}.
In order to map the phase diagram in the $\Theta^\star$--$T^\star$ plane, 
we have to set $K_{\rm c} = 1$ in (\ref{phb}) and solve for $T^\star_{\rm c}$ 
as a function of $\Theta^\star_{\rm c}$. 
This can be done numerically and the resulting phase diagram is shown on 
Fig.~\ref{fmf}.

\section{Results and Discussion}
\label{res}

To start the discussion, we note that the present study of the quantum APR model
was partly motivated by the strong changes
in shape of the orientational order parameter $\Phi$
as a function of temperature
as the rotational constant was increasing from
its classical value $\Theta^\star =0$, see Fig.~3 in Ref.~\cite{qapr1}.  
For small enough $\Theta^\star$ it was found that the order parameter
decays monotonously with increasing temperature similarly to the
classical case. 
This is qualitatively different for larger $\Theta^\star$,
where $\Phi (T^\star )$ becomes a non-monotonous function of temperature. 
In the present paper, we study this regime in much more detail 
and vary in addition to $\Theta^\star$ 
also the linear dimension $L$ of the lattice.

In Fig.~\ref{fop}, the order parameter vs. temperature curves are shown 
for different system sizes and for two representative $\Theta^\star$ choices,  
$\Theta^\star = 0.6109$ in (a) and 0.6982 in (b).
As can be seen from the data, in case (a) the order parameter
at low temperatures has a non-zero value and on heating first increases
up to a maximum and then decreases, which means that the system develops 
most order at intermediate 
temperatures. The results obtained for different system sizes indicate
that in the high temperature region the order
parameter scales to zero with increasing system size, whereas the maximum
value is system size independent. Therefore there is a long-range 
ordered phase at low temperatures present
in addition to the disordered high temperature phase. 
This orientational ordering that takes place with decreasing temperature
results in the limit $\Theta^\star \to 0$ in the 
classical APR transition and the well-defined herringbone ordered
ground state. 
This picture changes, however, as the quantum fluctuations
get more pronounced. When the value of $\Theta^\star$ is
increased further, the maximum value of the long-range order parameter 
drops considerably with increasing system size, as can be seen for 
$\Theta^\star = 0.6982$ from Fig.~\ref{fop}~(b). 
The order parameter at our lowest
temperatures is now strongly depressed
and also decreases further with increasing linear dimension of the system.
The finite-size effects in this intermediate $\Theta^\star$ range
are thus pronounced not only at the high-temperature wing
of the order parameter vs. temperature curve, 
but at {\it all} temperatures and in particular also close to 
the maximum of the order parameter. 
This is already a hint that the residual order in this range
of $\Theta^\star$--values might actually be a finite-size artifact.  
From the previous study~\cite{qapr1}
we know that a further increase in $\Theta^\star$ would suppress 
the ordering even more and finally lead to a situation with
vanishing order parameter at any temperature. 
All this motivated us to analyze the behavior of the
system further in order to see whether the intermediate-temperature maximum 
and low-temperature reduction of ordering, suggestive of a reentrant 
orientational melting, are really associated with the latter phenomenon.
In order to answer such question one has to employ 
finite-size scaling techniques which would allow to find out whether 
a particular phase transition is present, and if so, to locate it.

Before we start with the numerical finite-size scaling
analysis of the PIMC data, let us pause
for a moment and discuss the mean-field predictions from the previous section.
The resulting phase diagram in the $\Theta^\star$-$T^\star$ plane
is shown in Fig.~\ref{fmf} with the solid line.
We note that in the classical limiting case $\Theta^\star\to 0$
we find $T_{\rm c}^\star = 2$ which
agrees with earlier mean-field calculations~\cite{chata}.
Any crossing of the phase boundary in the mean-field phase
diagram Fig.~\ref{fmf} corresponds to a second order phase transition; 
note that it is believed~\cite{fo,marx1} that 
for $\Theta^\star =0$, the classical APR model undergoes
a fluctuation driven weak first order transition at $T^\star\approx 0.76$ 
for $L\to\infty$.
We see that at zero temperature there is a quantum phase transition 
at the value of ${\Theta^\star_c}^{MF} = 1$. 
The most interesting feature of the phase 
diagram is that there is a region of rotational constants ranging 
from $1$ to roughly $1.25$ for which the system is ordered at intermediate 
temperatures but disordered in the ground state (reentrance). 
The intuitive explanation of this
phenomenon is the following. At low temperatures, the individual rotors are
mostly in their totally rotationally symmetric ground state, which does not
possess quadrupolar moment and therefore cannot induce ordering via the
quadrupolar term. At intermediate temperatures, 
the excited states with non-zero
quadrupolar moment become populated and induce ordering which persists to
larger values of the rotational constant.
According to the mean-field theory,
reentrance takes place for a rather broad range of rotational constants 
$1 < \Theta^\star < 1.25$.
This corresponds to a decrease of the critical rotational
constant by roughly 20\% from its maximum value 
of about $1.25$ to the value of 1 at the ground state transition, and
represents a well pronounced feature.
Concerning the validity of this mean-field result,
it is known that the mean-field approximation sometimes tends to
overemphasize or even create reentrant behavior,
as pointed out in Ref.~\cite{stratt}.
On the other hand, reentrant behavior
has been experimentally observed in
solid HD under compression~\cite{silvera}. 
This three-dimensional system, however,
although consisting of diatomic molecules
interacting via approximate quadrupolar interactions, differs 
fundamentally from our model in spatial
dimensionality and structure of the lattice as well as in the dimensionality
of the order parameter.
In order to settle the question of reentrance in the 2D quantum APR model, 
we present now the results of numerical simulations and analysis techniques.  

We start with the results for potential, kinetic and total energies of our
model.
These are presented in Figs.~\ref{fe20} and \ref{fe23} for
two representative cases, $\Theta^\star = 0.6109$ and 
$\Theta^\star = 0.6982$, respectively.  
In both cases the potential energy first decreases with 
temperature until maximum order is achieved, while 
the kinetic energy increases strongly in this region.
For larger temperatures both $E_{\rm pot}$ and $E_{\rm kin}$ 
increase with temperature. 
Thus, when the temperature is increased from zero
the rotators occupy higher rotational states which allow for 
more pronounced orientational ordering
compared to that in the ground state. In these ordered states the 
attractive quadrupolar interaction is larger resulting in a lower potential
energy, and the kinetic energy increases strongly with temperature
due to enhanced localization of the rotators along some direction. 
This scenario continues with temperature until maximum order
is achieved. Further increase of temperature results in an increase
of $E_{\rm pot}$ due to thermal disorder making the quadrupolar interaction 
less important compared to the thermal energy. This can finally lead
to a phase transition from the low temperature orientationally ordered
phase to a high temperature disordered phase, provided the rotational constant
does not exceed a certain value. 
At very low temperatures the slope of the total energy
as a function of temperature, i.e., the 
specific heat, approaches zero, as expected for quantum systems.
We did not find a strong size dependency of $E_{\rm tot}(T^\star)$
and thus the specific heat behavior also does not seem to
depend on the system size.

Similar to the total energy, the average spread $R_\varphi$, 
which is a measure of the quantum mechanical delocalization
of the rotators,
is not dependent on system size as shown in Fig.~\ref{fmsq}.
As can be seen from its definition (\ref{mspread}) 
this quantity is a single particle property that is by construction not
particularly sensitive to collective effects.
The spread approaches its classical limit, i.e., zero, for
large temperatures, the approach being slower for larger $\Theta^\star$.  
In the limit of low temperatures, it reaches a ground state
value that is larger than 90$^\circ$ for the
rotational constant $\Theta^\star = 0.6982$, which means that
an individual rotator is no more confined to perform
librational motion around a preferred orientation but is instead
strongly delocalized.
On the other hand we know already from Fig.~\ref{fop}~(b)
that the order parameter is vanishingly small at low temperatures
and decreases even more with increasing system size for this
value of $\Theta^\star$.
The behavior of the average spread
is thus a clear demonstration that it is the quantum tunneling
which induces the disordering of the ground state for sufficiently
large rotational constants. 

In order to finally address the question whether our system 
has a reentrant phase transition as predicted by the mean-field study
we analyzed the low temperature region by the cumulant intersection
finite-size scaling method described in Sect.~\ref{finitesize},
see Figs.~\ref{fu4}, \ref{fu2}, and \ref{fu4th}. 
For our smallest rotational constant
$\Theta^\star=0.6109$  we clearly find 
an intersection point for 
both $U_L^{(4)}$ and $U_L^{(2)}$ at about $T^\star \approx 0.303$, 
see Figs.~\ref{fu4}(a)
and \ref{fu2}(a).
For the larger rotational constant
$\Theta^\star = 0.6364$ (b), the intersections
of the fourth and second order cumulants occur again both at the
same value within the error bars.
These crossings arise because the cumulants for larger
systems approach their limiting values faster than those for smaller systems.
For $\Theta^\star = 0.6666$ (c) the cumulants on the
different length scales
at low temperatures have values which cannot be distinguished
within the error bars, see Figs.~\ref{fu4}(c) and \ref{fu2}(c). 
The large value of $U^\ast \approx 0.65$ that was found for the 
classical APR model~\cite{fo} 
causes here the problem that within our numerical
accuracy we cannot identify an intersection point but rather obtain
a whole temperature region 
that is characterized by pronounced fluctuations. 
For larger $\Theta^\star$ constants the behavior 
of the two cumulants can again be distinguished fairly well.
However, contrary to what we found for $\Theta^\star \leq 0.6364$, 
the larger system has now the smaller cumulant throughout the
whole temperature range, see Figs.~\ref{fu4}(d) and \ref{fu2}(d). 
This signals the presence of the orientationally disordered phase
for $\Theta^\star=0.6982$ that extends from the ground state
up to high temperatures. From this behavior of the cumulants
we conclude that there is no evidence for the reentrance transition but rather
a pronounced increase of short-range order at intermediate temperatures
in the neighborhood of the phase boundary line. The latter has at low
temperatures a roughly vertical (i.e. $\Theta^\star$-independent) 
slope, rather than the characteristic shape suggested by the mean-field result 
in Fig.~\ref{fmf}.

In order to reinforce such conclusion for this part of the phase boundary  
we also studied the cumulants as functions of the rotational constant 
at constant temperature in the range $T^\star \leq 0.2424$. 
The fourth order cumulant plots of Fig.~\ref{fu4th} 
show that the intersection occurs in this low temperature range
at a value of about 
$\Theta^\star \approx 0.667 \pm 0.015$, 
which stays constant within the numerical noise;
the second order cumulants $U_L^{(2)}$ show the same behavior. 
Note that a systematic increase followed by a decrease of 
this quantity is expected as a function of temperature if
reentrance does occur.  
We can furthermore infer that the nontrivial value $U^\ast$ of the cumulants
at crossing is also in this low temperature regime with 
pronounced tunneling and quantum effects within the
numerical accuracy identical to the
value $U^\ast \approx 0.65$ found 
for the classical APR phase transition~\cite{fo}. 
We are thus lead to conclude that the APR transition temperature  
decreases slowly from its classical limit value of $T^\star \approx 0.76$ 
at $\Theta^\star = 0$ down to about $T^\star \approx 0.24$ at 
$\Theta^\star \approx 0.67$, where it suddenly drops dramatically. 
This numerically obtained non-reentrant phase
diagram is included in Fig.~\ref{fmf}. 
The PIMC simulation results clearly exclude the existence of a strongly  
pronounced reentrant feature in the phase diagram. Of course, we cannot exclude
the possible existence of a narrow reentrance region falling 
within the error bars of the present data. However, in any case these error bars
are much smaller than the 20 \% decrease
of the critical rotational constant predicted by the mean-field theory. 
The latter approximation is successful in predicting the phenomenon of enhanced
ordering at intermediate temperatures, 
but the deficiency is that it exaggerates the range of order and incorrectly 
predicts it to become long-ranged. As can be seen from Fig.~\ref{fmf},  
the mean-field theory apparently treats the quantum fluctuations 
(the limit $T^\star \to 0$)
better than the thermal fluctuations
(the limit $\Theta^\star \to 0$), 
since the transition point of the quantum induced transition
is overestimated only by a factor of about 1.5, whereas the
purely classical transition is off by a factor of more than 2.

\section{Summary and Conclusions}
\label{sum}
 
We have studied the quantum generalization of the anisotropic-planar-rotor model
consisting of point quadrupoles that are pinned with their
center of rotation on a triangular lattice. 
This two-dimensional statistical mechanical model system 
exhibits interesting orientational ordering effects, 
as a function of temperature and the rotational constant 
which controls the strength of the quantum fluctuations. 
A mean-field study predicts for this quantum system a reentrant 
orientational order-disorder transition
with a pronounced reentrance region.
Thus, in an intermediate range of rotational 
constant values at low temperatures the system orders with
increasing temperature, and then disorders again 
at higher temperatures via a phase transition, which corresponds
to the well-known herringbone orientational transition
in the classical limit.  

We studied the quantum APR model also numerically 
using path integral Monte Carlo simulations in
combination with a finite-size scaling analysis in order
to explore the region of the phase diagram where reentrance 
is expected.
It turned out that no reentrant transition is present, at least
within the usual numerical limitations of such simulations, 
which are most importantly set by the statistical errors and limited 
system sizes concerning both the $N$ and $P$ dimensionalities. 
In order to determine whether there still is a tiny reentrance region
hidden within our numerical accuracy, at least an order of magnitude larger
amount of computer time would be required, which at present seems to be 
prohibitively expensive.
For the same reason we did not attempt to address the issue of the 
order of the 
quantum induced APR phase transition in the ground state.  
We could, however, infer that the non-trivial cumulant values 
at the APR phase transition at low temperatures and large
rotational constants are within the estimated uncertainty 
identical to the value obtained in the classical limit.  

There are several possibilities to extend the present study.
Concerning the problem of reentrance in systems consisting
of quantum rotators, it might be interesting to study the
effect of higher multipole interactions.
In addition, 
it would be desirable to go beyond the mean-field approximation
by means of analytical techniques, such as, e.g., 
renormalization techniques capable of including accurately the
quantum fluctuations. 
Finally, it is certainly desirable to develop even more efficient
quantum simulation and analysis techniques.  
All this demonstrates that investigations of
{\em phase transitions} in {\em quantum} systems with continuous degrees 
of freedom are even today still a challenge.

\acknowledgements
 
RM acknowledges stimulating
discussions with E. Tosatti.
PN thanks the DFG for support (Heisenberg Foundation).
We gratefully acknowledge granted computer time on the
Cray--YMP (HLRZ J\"ulich and RHRK Kaiserslautern)
and CRAY--T90 (HLRZ J\"ulich).


 
\unitlength1mm 
\begin{figure}
\caption[]{
Orientational order parameter $\Phi$ as a function of temperature. 
Symbols indicate different system sizes, 
statistical error bars are shown, 
lines are for visual help only. 
The values of the rotational constants are: 
(a) $\Theta^\star = 0.6109$,
(b) $\Theta^\star = 0.6982$.
}
\label{fop}
\end{figure}
 
\begin{figure}
\caption[]{
Phase diagram of the quantum APR model in the 
$\Theta^\star$--$T^\star$ plane. 
The solid curve shows the line of continuous
phase transitions from an ordered phase at low temperatures
and small rotational constants to a disordered phase
according to the mean-field approximation.
The symbols show the transitions
found by the finite-size scaling analysis of the
path integral Monte Carlo data. 
The dashed line connecting these data is for visual help only.
}
\label{fmf}
\end{figure}
 
\begin{figure}
\caption[]{
Energies as a function of temperature for the rotational constant
$\Theta^\star = 0.6109$. 
(a) potential energy $E_{\rm pot}$,
(b) kinetic energy $E_{\rm kin}$,
(c) total energy $E_{\rm tot}$. 
Symbols indicate different system sizes, 
lines are for visual help only. 
}
\label{fe20}
\end{figure}
 
\begin{figure}
\caption[]{
Energies as a function of temperature for the rotational constant
$\Theta^\star = 0.6982$.
(a) potential energy $E_{\rm pot}$,
(b) kinetic energy $E_{\rm kin}$,
(c) total energy $E_{\rm tot}$.
Symbols indicate different system sizes, 
lines are for visual help only. 
}
\label{fe23}
\end{figure}

\begin{figure}
\caption[]{
Average spread $R_\varphi$ in degrees as a function of temperature. 
The values of the rotational constants are: 
(a) $\Theta^\star = 0.6109$,
(b) $\Theta^\star = 0.6982$.
Symbols indicate different system sizes, 
lines are for visual help only. 
}
\label{fmsq}
\end{figure}
 
\begin{figure}
\caption[]{
Fourth order cumulant $U_L^{(4)}$ as a function of temperature. 
The values of the rotational constants are: 
(a) $\Theta^\star = 0.6109$,
(b) $\Theta^\star = 0.6364$, 
(c) $\Theta^\star = 0.6667$, 
(d) $\Theta^\star = 0.6982$.
Symbols indicate different system sizes, 
lines are for visual help only. 
}
\label{fu4}
\end{figure}
 
\begin{figure}
\caption[]{
Second order cumulant $U_L^{(2)}$ as a function of temperature. 
The values of the rotational constants are: 
(a) $\Theta^\star = 0.6109$,
(b) $\Theta^\star = 0.6364$, 
(c) $\Theta^\star = 0.6667$, 
(d) $\Theta^\star = 0.6982$.
Symbols indicate different system sizes, 
lines are for visual help only. 
}
\label{fu2}
\end{figure}
 
\begin{figure}
\caption[]{
Fourth order cumulant $U_L^{(4)}$ as a function of rotational constant. 
The values of the temperature are: 
(a) $T^\star = 0.1515$.
(b) $T^\star = 0.1818$, 
(c) $T^\star = 0.2121$, 
(d) $T^\star = 0.2424$,
Symbols indicate different system sizes, 
lines are for visual help only. 
}
\label{fu4th}
\end{figure}
 
\end{document}